\documentclass[preprint]{elsarticle}
\usepackage{amssymb}
\usepackage[final]{epsfig}
\usepackage{color}
\usepackage{graphicx}
\usepackage{lineno}
\newcommand{\beq}{\begin{equation}}
\newcommand{\eeq}{\end{equation}}
\newcommand{\bdi}{\begin{displaymath}}
\newcommand{\edi}{\end{displaymath}}
\newcommand{\no}{\nonumber}
\newcommand{\bea}{\begin{eqnarray}}
\newcommand{\eea}{\end{eqnarray}}

\newcommand{\text}{}

\newcommand{\de}{\partial}

\begin{document}

\begin{frontmatter}

\title{An extension of the Gluckstern formulas for multiple scattering: analytic expressions for track parameter resolution using optimum weights}

\author[add1,add2]{Z. Drasal} 
\author[add2]{W. Riegler \corref{cor1}}
\cortext[cor1]{Corresponding author: werner.riegler@cern.ch} 
\address[add1]{Charles University, Prague, Czech Republic}
\address[add2]{CERN EP, Geneva, Switzerland}


\begin{abstract}
         Momentum, track angle and impact parameter resolution are key performance parameters that tracking detectors are optimised for. This report presents analytic expressions for the resolution of these parameters for equal and equidistant tracking layers. The expressions for the contribution from position resolution are based on the Gluckstern formulas and are well established. The expressions for the contribution from multiple scattering using optimum weights are discussed in detail.            
\end{abstract}

\end{frontmatter} 

\section{Introduction}

The theory of track fitting using global $\chi^2$ minimisation is well established \cite{rudi} \cite{mankel} and some explicit expressions for geometries with equidistant detector planes are presented in \cite{gluckstern} \cite{rudi_meinhard1} \cite{rudi_meinhard2}. In this report we derive analytic expressions for the resolution of particle momentum as well as track angle and impact parameter in $r{-}\phi$ and $z$ direction, as defined in Fig. \ref{rphiz}. The calculations are performed for a classic solenoid spectrometer with a constant B-field using $N+1$ equal and equidistant detector planes. We present both, the contribution from detector resolution and the contribution from multiple scattering for each of these 5 parameters. In the following, we first present the formalism for $\chi^2$ minimisation, then we calculate the covariance matrix of individual measurements, and it's inverse, assuming detector position resolution and multiple scattering. Then we derive the covariance matrix for the parameters of a straight line track and a parabolic track in an $x{-}y$ coordinate system and finally, we use these results to write down the errors on track parameters in the summary section.

\clearpage
\newpage

\section{General formulas}

We assume a particle track of known functional form $f(x) = \sum_{i=0}^M a_i g_i(x)$ with $M+1$ unknown parameters $a_m$, and we assume $y_n$ to be the measured positions in the $N+1$ detector planes positioned at $x_n$. The straight line track in Fig. \ref{straight_line} and the parabolic track in Fig. \ref{parabola} are the two concrete examples that we will discuss later. The parameters $a_i$ are estimated by minimising $\chi^2$ defined as  
\beq
     \chi^2 = \sum_{m=0}^N \sum_{n=0}^N 
     \left[
     y_m-\sum_{i=0}^M a_ig_i(x_m)
     \right]
     \,W_{mn} \,
     \left[
     y_n - \sum_{i=0}^M a_ig_i(x_n)
     \right]
\eeq
where $W_{mn}$ is the weight matrix that still has to be defined. The above relation can also be written in matrix form
\beq
    \chi^2 = ({\bf y-  G a})^T{ \bf W( y-Ga) }
\eeq
with ${\bf a} = (a_0, a_1, ..., a_M)$, ${\bf y} = ( y_0, y_1, ..., y_N)$ and $G_{mn} = g_n(x_m)$. To minimise $\chi^2$ we have to solve $\de \chi^2 / \de a_i = 0$ which gives
\beq
    {\bf  a} =({\bf G}^T {\bf WG})^{-1} {\bf G}^T {\bf W\,y}= {\bf B\,y}
\eeq
and represents the estimates for the parameters $a_i$. Next we want to know the variance of these estimated parameters for given measurement errors on $y_n$. These errors are defined through the covariance matrix ${\bf C_y}$ of ${\bf y}$. From error propagation we know that if ${\bf C}_y$ is the covariance matrix for ${\bf y}$, the covariance matrix ${\bf C}_a$ for ${\bf a} = {\bf B y}$  is 
\beq
    {\bf   C}_a = {\bf B} {\bf C}_y {\bf B}^T= ({\bf G}^T{\bf WG})^{-1} {\bf G}^T {\bf W} \, {\bf  C}_y\, {\bf W}^T {\bf G} ({\bf G}^T{\bf W}^T{\bf G})^{-1} 
\eeq
which is the desired result. The variance of the track position $f(x)$ and track angle $f'(x)$ along the track are then given in analogy by
\beq \label{covariance_along_track}
    (\Delta f(x))^2 = {\bf g}(x)^T\,{\bf C}_a {\bf g}(x) \qquad \qquad  (\Delta f'(x))^2 = {\bf g}'(x)^T\,{\bf C}_a {\bf g}'(x) 
\eeq
with ${\bf g}(x) = (g_0(x), g_1(x), ..., g_M(x))$ and ${\bf g}'(x) = (g_0'(x), g_1'(x), ..., g_M'(x))$. The weight matrix ${\bf W}$ has to be chosen such that the variances are minimised and the estimators are unbiased.  This question is answered by the generalized Gauss-Markov theorem, stating that ${\bf W}={\bf C}_y^{-1}$ is the optimum choice. In that case the expression for ${\bf C}_a$ reduces to
\beq \label{optimum_variance}
      {\bf  C}_a  =  ({\bf G}^T {\bf C}_ y^{-1} {\bf G} )^{-1} 
\eeq

\clearpage
\newpage

\section{Covariance matrix $C_y$}

\begin{figure}[ht]
 \begin{center}
   \includegraphics[width=10cm]{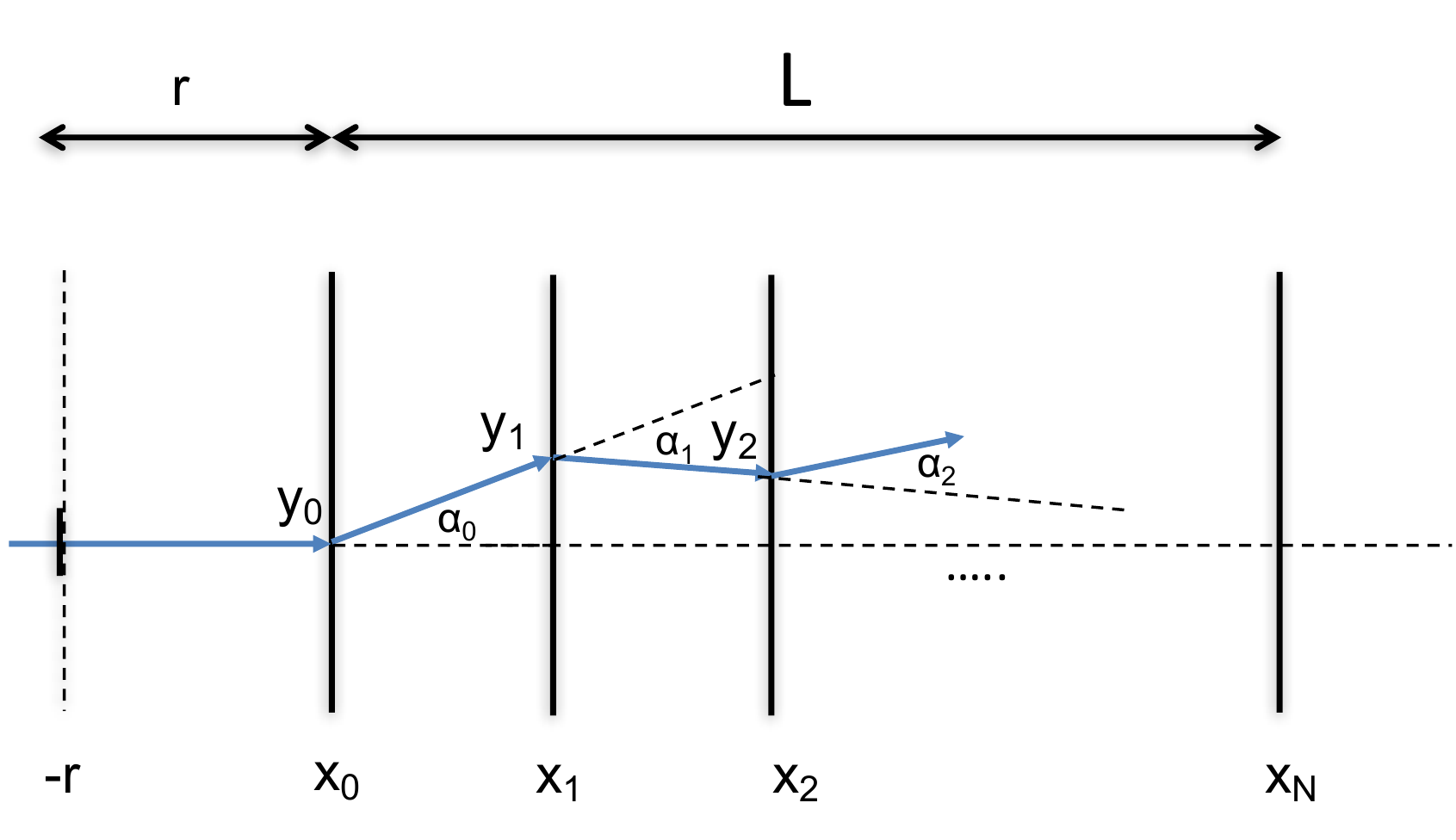}
  \caption{Effect of multiple scattering in the different detector planes.}
  \label{multiple_scattering}
  \end{center}
\end{figure}

There are two sources for the measurement errors on $y_n$, the position resolution $\sigma_i$ of the detector planes, which are uncorrelated, and multiple scattering in the detector planes, which are highly correlated. When assuming thin scatterers, the variance  $\sigma_{\alpha_i}$ of the multiple scattering angle in a single detector plane is given by \cite{pdg}
\beq
     \sigma_{\alpha_i} = \frac{0.0136\,\mathrm{GeV/c}}{\beta p}\,\sqrt{\frac{d_i}{X_0}}\left(1+0.038 \ln \frac{d_i}{X_0}\right) = 
    \frac{1}{\beta p[GeV/c]}\,\ f \left( \frac{d_i}{X_0}\right) 
\eeq
where $d_i/X_0$ is the thickness of a single detector plane in units of radiation length and 
\beq
f(y)=0.0136\,\mathrm{GeV/c}\sqrt{y}(1+0.038\ln y).
\eeq
Fig.  \ref{multiple_scattering} shows how these errors affect the measurements in the different planes.
Assuming that 'for a single event' we have an offset $u_n$, with a mean value of zero and variance $\sigma_n^2$ due to detector resolution, and a scattering angle $\alpha_n$ in the $n^{th}$ detector plane, the measurement values $y_n$ are
\bea
      y_0 &=&  f(x_0) + u_0   \no \\
      y_1 &=&  f(x_1)+u_1+\alpha_0(x_1-x_0) \no \\
      y_2 &=&  f(x_2)+ u_2+\alpha_0(x_2-x_0)+\alpha_1(x_2-x_1) \no \\
      \vdots &&  \no \\
      y_n &=& f(x_n)+u_n+\sum_{m=0}^{n-1} \alpha_m(x_n-x_m) \qquad n=0, 1, ..., N \no
\eea
The covariance matrix of $y_n$ is therefore  
\beq
   (C_y)_{mn} = \sigma_n^2\delta_{mn}+\sum_{j=0}^{\mathrm{Min}[m,n]-1}\,\sigma_{\alpha_j}^2\,(x_m-x_j)(x_n-x_j)
\eeq
In case all detector planes have the same position resolution ($\sigma_i^2=\sigma^2$), the planes are equidistant ($x_n=nL/N$, $n=0,1, ..., N$) and the detector planes have identical material budget i.e. identical multiple scattering effect ($\sigma_{\alpha_j}=\sigma_\alpha$), we have 
\bea
   (C_y)_{mn}  =  \sigma^2\delta_{mn}+\frac{\sigma_{\alpha}^2}{6}\,\left(\frac{L}{N}\right)^2\, n(3m+3mn+1-n^2)  \quad m \ge n 
    \quad  \mbox{with} \quad  M_{nm} = M_{mn}
\eea
This matrix is used to find the covariance matrix ${\bf C}_a$ for the combined effect from position resolution and multiple scattering through Eq. \ref{optimum_variance}. In order to be able to derive some elementary formulas we investigate two limiting cases where either the detector resolution dominates or the multiple scattering dominates. In case the detector resolution plays the dominant role we set $\sigma_{\alpha}=0$ and have 
\beq
    {\bf C}_y ={\bf  R} = \sigma^2 {\bf 1} \qquad   {\bf C}_y^{-1} = { \bf R}^{-1} = \frac{1}{\sigma^2} {\bf 1}
\eeq
In case multiple scattering dominates we set $\sigma=0$, and the covariance matrix explicitly reads as
\beq
      {\bf C}_y ={\bf M} = 
\frac{\sigma_{\alpha}^2 L^2}{N^2}
\left(
\begin{array}{ccccccccccc}
 \frac{N^2 \sigma_0^2}{\sigma_{\alpha}^2 L^2} & 0 & 0 & 0 & 0 & 0 & 0 & 0 & . & . & . \\
 0 & 1 & 2 & 3 & 4 & 5 & 6 & 7 &  \ldots \\
 0 & 2 & 5 & 8 & 11 & 14 & 17 & 20 & \ldots \\
 0 & 3 & 8 & 14 & 20 & 26 & 32 & 38 & \ldots \\
 0 & 4 & 11 & 20 & 30 & 40 & 50 & 60 & \ldots \\
 0 & 5 & 14 & 26 & 40 & 55 & 70 & 85 & \ldots \\
 0 & 6 & 17 & 32 & 50 & 70 & 91 & 112 & \ldots \\
 0 & 7 & 20 & 38 & 60 & 85 & 112 & 140 & \ldots \\
 \vdots & \vdots &\vdots & \vdots & \vdots & \vdots & \vdots & \vdots & \ddots
\end{array}
\right)
\eeq
In order to avoid a singular matrix we still keep $\sigma_0$ finite and take the limit to zero only for the final result. The inverse of this matrix can be calculated explicitly for every $N$ and is given by
\beq \label{m_inverse}
   {\bf C}_y^{-1}= {\bf M}^{-1} \, = \, 
    \frac{N^2}{\sigma_{\alpha}^2L^2} \left(
\begin{array}{cccccccccccccc}
 \frac{L^2 \sigma_{\alpha}^2}{N^2\text{\sigma_0}^2 } & 0 & 0 & 0 & . & . & . & . & . & . & .&  . \\
 0 & 6 & -4 & 1 & 0 & . & . & . & . & . & . & .\\
 0 & -4 & 6 & -4 & 1 & 0 & . & .& . & . & . & .\\
 0 & 1 & -4 & 6 & -4 & 1 & 0 & . & . & . & . & .\\
 . & 0 & 1 & -4 & 6 & -4 & 1 & 0 & . & . & .  & .\\
 . & . & . & . & . & . & . & .& . & . & . & .  \\
. & . & . & . & . & . & . & .& . & . & .  & . \\
. & . & . & . & . & . & . & .& . & . & .  & . \\
   . & . & . & . & .  & 0 & 1 & -4 & 6 & -4 & 1 & 0  \\
  . & . & . &. & . & .  & 0 & 1 & -4 & 6 & -4 & 1 \\
   . & . & . &. & . & .  & . & 0 & 1 & -4 & 5 & -2  \\
   . & . & . &. & . & . &.  & . & 0 & 1 & -2 & 1 \\
\end{array}
\right)
\eeq

\clearpage
\newpage

\section{Straight line track}

\begin{figure}[ht]
 \begin{center}
   \includegraphics[width=10cm]{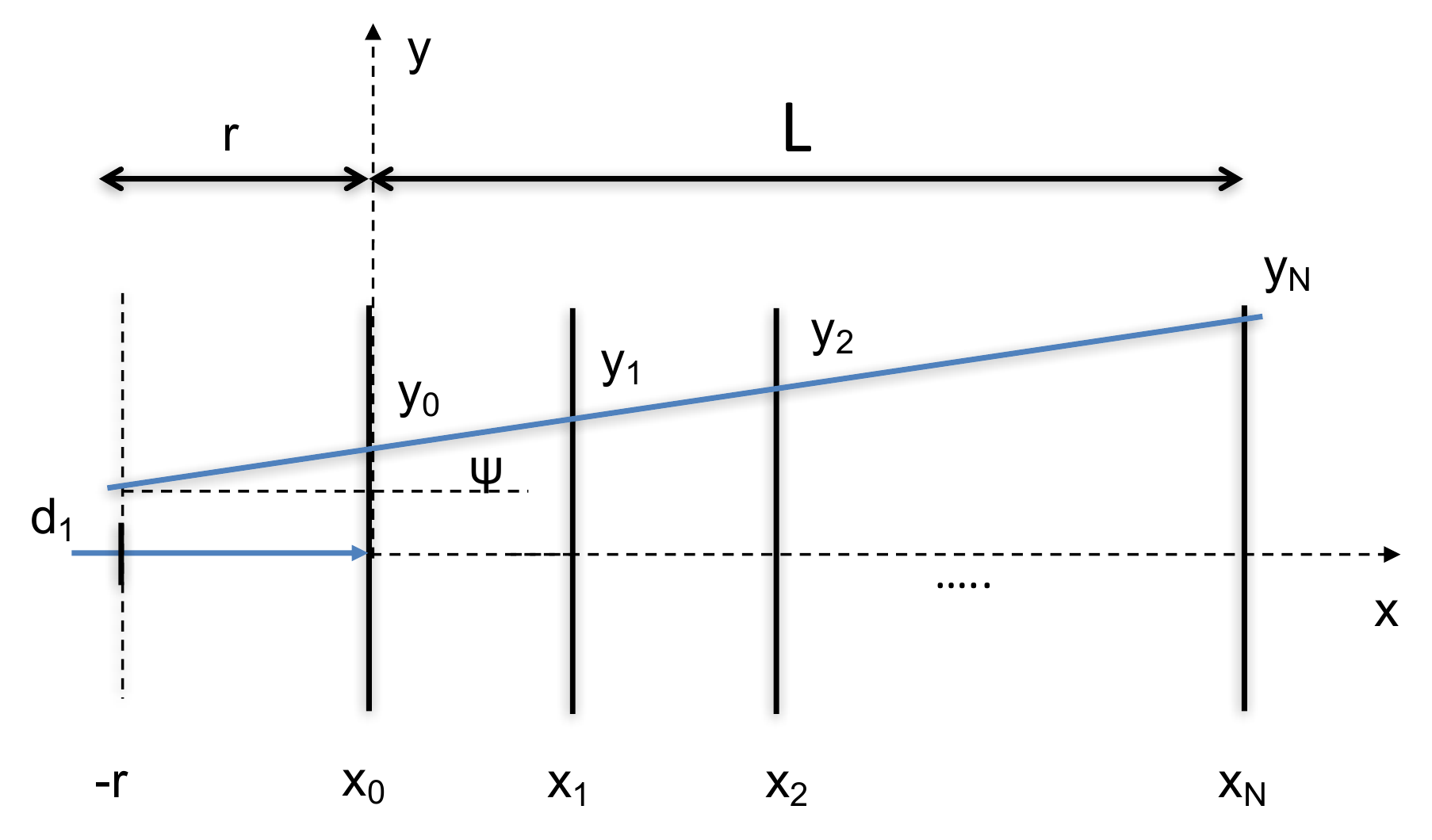}
  \caption{A straight line track through $N+1$ equal and equidistant detector planes.}
  \label{straight_line}
  \end{center}
\end{figure}
\noindent
We assume the geometry shown in Fig. \ref{straight_line} with a straight line track $f(x)=a_0+a_1x$ through $N+1$ equal and equidistant detector planes. We also assume the track to be almost parallel to the $x$-axis such that $\tan \psi \approx \psi \approx f'(x)=a_1$ and treat larger track inclinations later in the summary. We have $g_0=1$ and $g_1=x$, and with $x_n=nL/N$ we get 
\beq
  {\bf  G}^T = \left(
    \begin{array}{cccccc}
        1& 1& 1& 1&  ... & 1 \\
         0 & \frac{L}{N}& \frac{2L}{N}&  \frac{3L}{N} & ... & L \\       
    \end{array}
    \right)
\eeq
For the contribution from detector resolution we have ${\bf C}_y = {\bf R}$ and therefore (c.f. Eq. 25 in \cite{gluckstern} )
\beq
   {\bf  C}_a = ({\bf G}^T {\bf R}^{-1}{\bf G})^{-1} = \sigma^2  ({\bf G}^T {\bf G})^{-1} = \frac{\sigma^2}{(N+1)(N+2)}
    \left(
       \begin{array}{cc}
         2(2N+1) & -\frac{6N}{L} \\
         -\frac{6N}{L}  & \frac{12N}{L^2}
       \end{array}
    \right)
\eeq
The variance of the track angle $\Delta \psi = \Delta a_1=\sqrt{(C_a)_{11}}$ is given by
\beq \label{psi_reso}
     \Delta \psi = \frac{\sigma}{L}\sqrt{\frac{12N}{(N+1)(N+2)}}
\eeq
To find the 'impact parameter' $d_1$ we have $d_1=f(-r)$ and with Eq. \ref{covariance_along_track} we find the variance of  $d_1$ as
\bea \label{delta_z0_res}
       (\Delta d_1)^2 =(\Delta f(-r))^2
       & = & \frac{4\sigma^2}{(N+1)(N+2)}
       \left[ \left(N+\frac{1}{2} \right)+\frac{3N r}{L}+\frac{3N r^2}{L^2} \right] \\
       & \approx & 
       \frac{4\sigma^2}{N+3}
         \left(1+\frac{3 r}{L}+\frac{3 r^2}{L^2} \right) \quad \mbox{for}  \quad N \gg 1
\eea
For very small values of $r/L$ we have $\Delta d_1 = \sigma/0.91\sigma/0.84\sigma/0.77\sigma$ for $N=1/2/3/4$. For $r/L=1/10$ we have $\Delta d_1 = 1.1\sigma/1.0 \sigma/0.95 \sigma/0.88\sigma$.

Next we consider the contribution due to multiple scattering, where we first use equal weights for all measurement points to illustrate the difference to optimum weights. We use ${\bf W}={\bf 1}$ and get
\beq
    {\bf  C}_a  = ({\bf G}^T{\bf G})^{-1}({\bf G}^T \,{\bf M\, G}) ({\bf G}^T{\bf G})^{-1}  = \frac{\sigma_\alpha^2}{210N(N+1)(N+2)} \times
\eeq
\beq
   \left(
     \begin{array}{cc}
      L^2(2N^4+N^3-2N-1) & -L(11N^4+16N^3-14N^2-25N+12) \\[2mm]
      -L(11N^4+16N^3-14N^2-25N+12)  & 78N^4+312N^3+462N^2+300N+108
     \end{array}
     \right)
\eeq
The variance of $d_1$ is  
 \beq
 \begin{array}{cclc}
        (\Delta d_1)^2 =(\Delta f(-r))^2
        & = & \sigma_{\alpha}^2 r^2 & N=1 \\[2mm]
        & = &  \sigma_{\alpha}^2 r^2 (5/4+L/12r+L^2/144r^2) & N=2 \\[2mm]
        &  & \vdots & \\[2mm]
         & \approx &  N \sigma_{\alpha}^2 r^2 (39/105+11L/105r+L^2/105r^2) & N \gg 1
        \end{array}
 \eeq
For $N=1$, i.e for two layers, the $d_1$ resolution is $\sigma_\alpha r$ and it deteriorates as one introduces more layers, so the equal weights are clearly not ideal for the best measurement precision. Using the optimum weights for the multiple scattering limit i.e. ${\bf W}={\bf M}^{-1}$ from Eq. \ref{m_inverse}, we find
 \beq
  {\bf   C}_a = \lim_{\sigma_0 \rightarrow 0}({\bf G}^T {\bf M}^{-1} {\bf G})^{-1} = 
     \left(
     \begin{array}{cc}
    0 &0 \\[2mm]
      0  & \sigma_{\alpha}^2
     \end{array}
     \right)
\eeq
There is no dependence on $N$, so the angular resolution is independent of the number of detector layers 
\beq \label{psi_ms}
      \Delta \psi = \sigma_{\alpha}
\eeq
and equal to the scattering error in a single detector layer. One therefore does not improve the resolution by adding more detector layers, because the additional measurement information is 'cancelled' by the additional scattering in the detector material. We'll see later that this holds only for the straight line fit. For the $d_1$ resolution we have
\beq \label{delta_z0_ms}
   \Delta d_1 = r \, \sigma_{\alpha}
\eeq
which is also independent of the number of detector layers.

\clearpage
\newpage

\section{Parabolic track}

\begin{figure}[ht]
 \begin{center}
   \includegraphics[width=10cm]{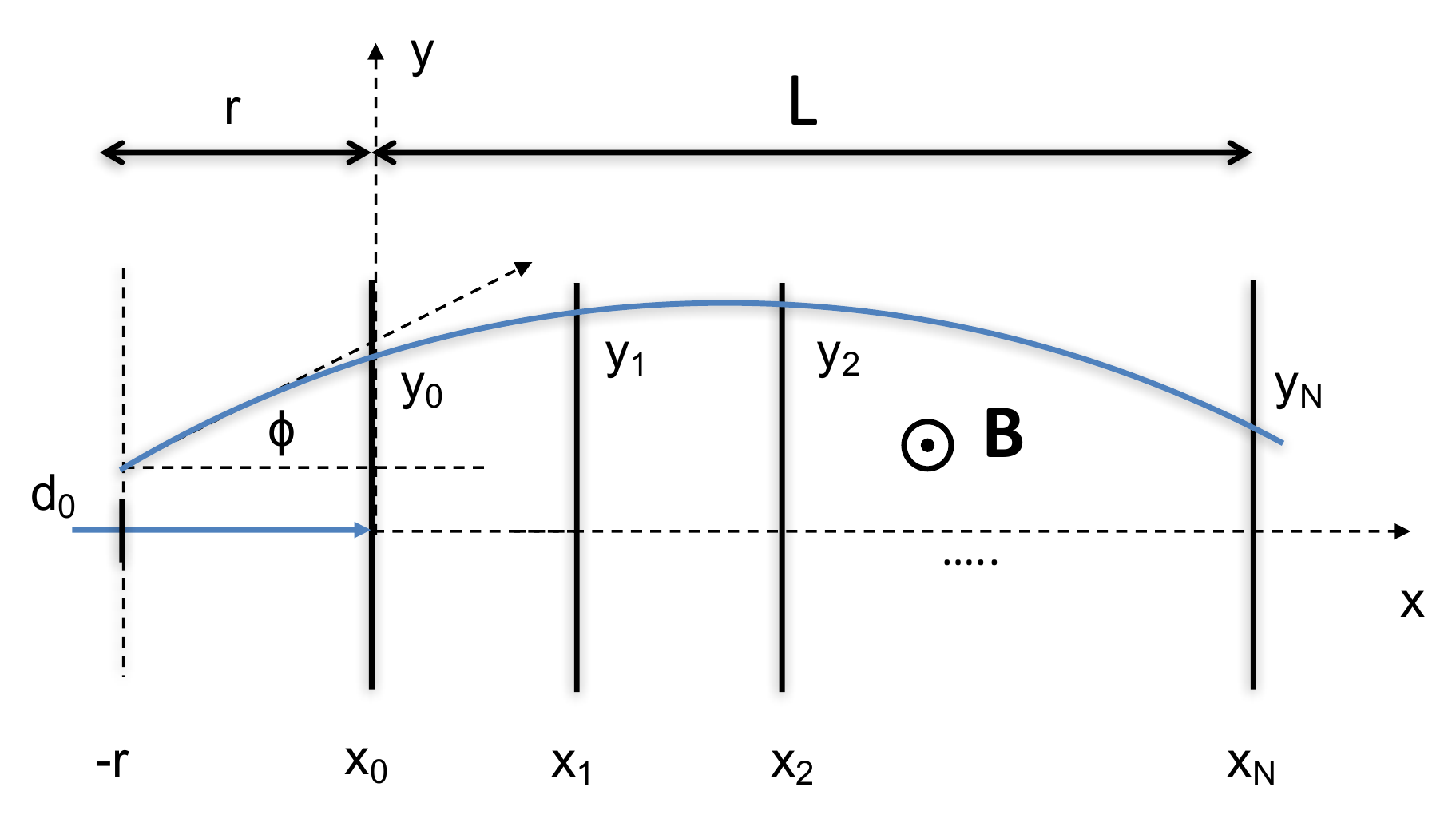}
  \caption{A parabolic track through $N+1$ equal and equidistant detector planes.}
  \label{parabola}
  \end{center}
\end{figure}
\noindent

We assume the geometry shown in Fig. \ref{parabola} where a particle of momentum $p$ describes a circle of radius $R$[m]=$p$[GeV/c]/(0.3$B$[T]) in the magnetic field. We approximate this circle by $f(x)=a_0+a_1x+a_2x^2/2$ with $a_2=1/R$, such that the momentum resolution becomes
\beq \label{momentum_resolution_form1}
        \frac{\Delta p}{p} = \frac{p}{0.3 B}\, \Delta a_2 
\eeq
As for the straight line track we assume $a_1$ to be small such that $\tan \phi \approx \phi \approx f'(x)=a_1+a_2x$ along the track. We have $g_0=1, g_1=x, g_2=x^2/2$ and therefore 
\beq
 {\bf   G}^T = \left(
    \begin{array}{cccccc}
        1& 1& 1& 1 & ... & 1 \\
         0 & \frac{L}{N}& \frac{2L}{N}& \frac{3L}{N}&... & L \\       
                  0 & \frac{1}{2}\left(\frac{L}{N}\right)^2 & \frac{1}{2}\left(\frac{2L}{N}\right)^2 & \frac{1}{2}\left(\frac{3L}{N}\right)^2& ... & \frac{1}{2}L^2       
    \end{array}
    \right)
\eeq
and the covariance matrix for the case where the detector resolution $\sigma$ dominates over multiple scattering is (c.f. Eq. 13 in \cite{gluckstern})
\beq \label{covariance_parabola}
  {\bf  C}_a = ({\bf G}^T {\bf R}^{-1} {\bf G})^{-1}  = 
 \frac{\sigma^2}{(N+1)(N+2)(N+3)}
\left(
\begin{array}{ccc}
 9 N (N+1)+6 & -\frac{18 N (2 N+1)}{L} & \frac{60 N^2}{L^2} \\
 -\frac{18 N (2 N+1)}{L} & \frac{12 N (2 N+1) (8 N-3)}{L^2 (N-1)} & -\frac{360 N^3}{L^3 (N-1)} \\
 \frac{60 N^2}{L^2} & -\frac{360 N^3}{L^3 (N-1)} & \frac{720 N^3}{L^4 (N-1)} \\
\end{array}
\right)
\eeq
The momentum resolution is therefore
\beq \label{momentum_resolution_raw}
            \frac{\Delta p}{p} = \frac{p}{0.3 B}\, \sqrt{(C_a)_{22}} = \frac{\sigma \, p}{0.3 B\,L^2}
           \sqrt{ \frac{720 N^3}{(N-1)(N+1)(N+2)(N+3)}}
\eeq
The variance on the track angle $\phi$ reads as
\beq \label{phi_angle_reso}
      (\Delta f'(-r))^2 = \frac{12 \sigma^2}{L^2 (N-1)(N+1)(N+2)(N+3)}
      \left(
      (16N^3+2N^2-3N)+\frac{60N^3\,r}{L}+\frac{60N^3\,r^2}{L^2}
      \right)
\eeq
The $d_0$ resolution is explicitly written in Eq. \ref{sum_d0_res} and for large values of $N$ it is approximated by 
\bea \label{delta_d0_res}
             (\Delta d_0)^2  = (\Delta f(-r))^2 
       & \approx&    \frac{9 \sigma^2}{N+5} \left(
        1+\frac{8r}{L} + \frac{28r^2}{L^2}+\frac{40 r^3}{L^3} + \frac{20 r^4}{L^4}
    \right)
\eea
For very small values of $r/L$ Eq. \ref{sum_d0_res} gives $\Delta d_0 = \sigma/0.97\sigma/0.94\sigma/0.91\sigma$ for $N=2/3/4/5$.  We see that the $d_1$ resolution for 2 layers is the same as the $d_0$ resolution for 3 layers, and for larger values of $N$ the $d_0$ resolution is always worse and approaches a ratio of $\sqrt{9/4}=1.5$ for large values of $N$. This reflects the fact that for the parabola there are 3 degrees of freedom while for the straight line there are only two. For $r/L=1/10$ Eq. \ref{sum_d0_res} gives $\Delta d_0 = 1.4\sigma/1.37\sigma/1.34\sigma/1.29\sigma$ for $N=2/3/4/5$,  significantly worse than $\Delta d_1$ from the straight line track. 
\\
For the situation where multiple scattering dominates we first apply equal weights in order to make the link to the results in \cite{gluckstern} and to specifically see the difference to optimum weights for the momentum resolution. With ${\bf W} {=} {\bf 1}$ we have 
\beq
    {\bf   C}_a    =  ({\bf G}^T {\bf G})^{-1}( {\bf G}^T {\bf M} {\bf G}) ({\bf G}^T{\bf G})^{-1}   
\eeq
and just quote the following elements from this matrix:
\bea
     (C_a)_{11} & = & \sigma_{\alpha}^2 \,  \frac{16 N^6+81 N^5+234 N^4+321 N^3+284 N^2-228 N-108}{70 (N-1) (N+1)(N+2)(N+3)N} 
                             \approx  \sigma_{\alpha}^2 N \frac{8}{35} \\
     (C_a)_{12} & = &    - \sigma_{\alpha}^2 \, \frac{N \left(3 N^4+5N^3+15 N^2+55 N+162\right)}{14 L (N-1)(N+1)(N+2)(N+3)} 
       \approx  -\frac{\sigma_{\alpha}^2 N}{L} \frac{3}{14} \\
     (C_a)_{22} & = &    \sigma_{\alpha}^2 \, \frac{10 N \left(N^4+4 N^3+5 N^2+2 N+12\right)}{7 L^2 (N-1)(N+1)(N+2)(N+3) }  
       \approx  \frac{\sigma_{\alpha}^2 N}{L^2} \frac{10}{7} 
\eea
The limits of $10/7, 3/14, 8/35$ for large values of $N$ represent the limits of $C_N, D_N, E_N$ in Table 2 of \cite{gluckstern}. The momentum resolution for a large number of detector planes therefore becomes
\beq
   \frac{\Delta p}{p} = \frac{p}{0.3B} \sqrt{   (C_a)_{22} }
   \approx \sqrt{\frac{10}{7}}\,\frac{p\sigma_\alpha  \, \sqrt{N}}{0.3BL}  \approx 1.20\,\frac{p\sigma_\alpha  \, \sqrt{N}}{0.3BL}  \qquad N \gg 1
\eeq
It is quoted in \cite{scott} and \cite{gluckstern} that this factor 1.20 can be turned into unity in the limit of large $N$ for optimum weights, and a numerical evaluation for finite $N$ is given in \cite{rudi}. Using the optimum weight matrix ${\bf W} = {\bf M}^{-1}$ we can derive an explicit expression for  $(C_a)_{22}$. The covariance matrix is
\beq
  {\bf  C}_a = \lim_{\sigma_0 \rightarrow 0}({\bf G}^T {\bf M}^{-1} {\bf G} )^{-1}  = \sigma_\alpha^2 \, 
        \left(
    \begin{tabular}{ccc}
         0 & 0 & 0 \\
          0 & $\frac{N-3/4}{N-1}$ & $- \frac{N}{2(N-1)L}$ \\
          0 & $- \frac{N}{2(N-1)L}$ & $\frac{N^2}{(N-1)L^2}$
    \end{tabular}
    \right)
\eeq
The contribution of multiple scattering to the momentum resolution is therefore
\bea
     \frac{\Delta {p}}{p}   & = &   \frac{p}{0.3B} \sqrt{(C_a)_{22}}                  \\
     & = &\frac{N}{\sqrt{(N+1)(N-1)}} \, \frac{p\,\sigma_{\alpha} \sqrt{N+1}}{0.3BL} \\
     & =  &  (1.15, 1.06, 1.03, 1.02, ...) \, \frac{p\,\sigma_{\alpha} \sqrt{N+1}}{0.3BL} \qquad N=2, 3, 4, 5, ...
\eea
So the factor becomes indeed unity for large $N$ and the convergence is rather fast. Inserting the expression for $\sigma_\alpha$ we find 
\beq \label{multiple_scattering_final}
     \frac{\Delta {p}}{p}   =  \frac{N}{\sqrt{(N+1)(N-1)}} \, \frac{0.0136\,\mathrm{GeV/c}}{0.3 \beta B L }\,\sqrt{\frac{d_{tot}}{X_0}}\left(1+0.038 \ln \frac{d}{X_0}\right)
\eeq
where $d_{tot}= (N+1)d$ is the total thickness of all detector layers. We see that the contribution to the momentum resolution from multiple scattering is independent on the particle momentum $p$, and is mainly affected by the total material budget. The exception is for small momenta where $\beta = (p/\sqrt{m^2c^2+p^2)}$ is different from unity the resolution deteriorates accordingly.
\\ 
For the resolution of the angle $\phi$ we have
\beq \label{phi_angle_ms}
   \Delta \phi = \Delta f'(-r)  = \sigma_{\alpha} \,
   \sqrt{
   \frac{N-3/4}{N-1} 
   + \frac{N}{N-1}\left( \frac{r}{L} \right)
   +  \frac{N^2}{N-1}\left( \frac{r}{L} \right)^2
   } 
\eeq
While for angle of the straight line fit we have $\Delta \psi = \sigma_{\alpha}$ independent on the number of layers, $\Delta \phi$ is larger than $\sigma_\alpha$ and shows a dependence on the number of layers. The reason is related to the fact that for the parabola there are 3 instead of 2 degrees of freedom, so the track is less constrained. For 3 layers, i.e. $N{=}2$ and $r/L=1/10$ we have $\Delta \phi {=} 1.22\,\sigma_\alpha$ i.e. a 22\% worse resolution as compared to $\Delta \psi$. The expression actually has a minimum at $N=2+L/2r$ that evaluates to
\beq
    \Delta \phi = \sigma_\alpha \sqrt{1+\frac{2r}{L}+\frac{4r^2}{L^2}}
\eeq 
This means that given an allocated envelope $L$ for the tracker, there is an optimum number of layers inside this envelope that achieves the best possible $\phi$ resolution when considering multiple scattering only. For $r/L=1/10$ the best achievable resolution is $\Delta \phi {=} 1.11\,\sigma_\alpha$, so around 11\,\% worse that the $\Delta \psi$ resolution.
\\
The $d_0$ resolution is given by the variance of $f(-r)$ and reads as
\bea \label{delta_d0_ms}
     (  \Delta d_0)^2=  (\Delta f(-r))^2 & = & \sigma_{\alpha}^2 r^2 \left[
         \frac{N-3/4}{N-1} + 
         \frac{N}{2(N-1)} \left( \frac{r}{L} \right) + 
         \frac{N^2}{4(N-1)} \left( \frac{r}{L}\right)^2
      \right]
\eea
This $d_0$ resolution also has a minimum at $N=2+L/r$, different from the minimum for the $\phi$ resolution, where it evaluates to
\beq
       (\Delta d_0)^2 =\sigma_{\alpha}^2r^2 \left( 1 + \frac{r}{L} + \frac{r^2}{L^2}\right)
\eeq
In typical vertex detector layout, $r\ll L$, and hence $\Delta d_0 \approx \sigma_\alpha r$.
By assuming e.g. the first layer at r=2\,cm and a radial extent of the vertex tracker of $L=20$\,cm we have $L/r=10$ and the optimum $d_0$ resolution would be achieved with 13 layers and evaluate to $\Delta d_0 = 1.05 \sigma_{\alpha} r$, so only 5\% worse than the best possible resolution. For $N=3$ i.e. 4 layers the resolution is  $\Delta d_0 = 1.1 \sigma_{\alpha} r  $, so only 10\,\% worse than the limit case. \\
If we assume the distance between layers to be fixed to $D$ and consider adding more and more layers, we have $L=D N$ and the $d_0$ resolution approaches $\Delta d_0 = \sigma_\alpha r$ for large numbers of $N$.

\clearpage
\newpage



%
%
%
%
%
%
%
%
%
%
%
%

\clearpage
\newpage

\section{Summary}

\begin{figure}[ht]
 \begin{center}
 a)
   \includegraphics[width=6.5cm]{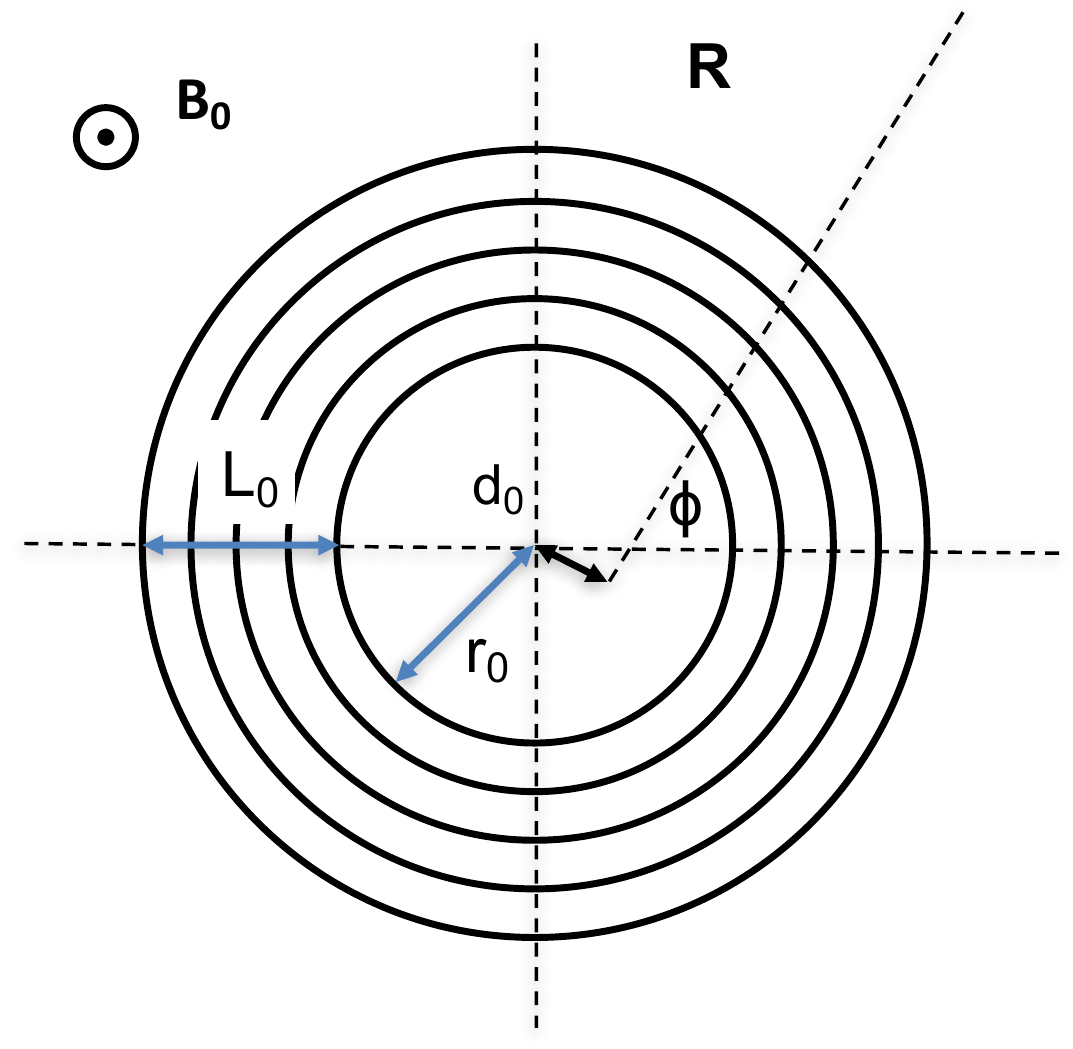}
   b)
   \includegraphics[width=7cm]{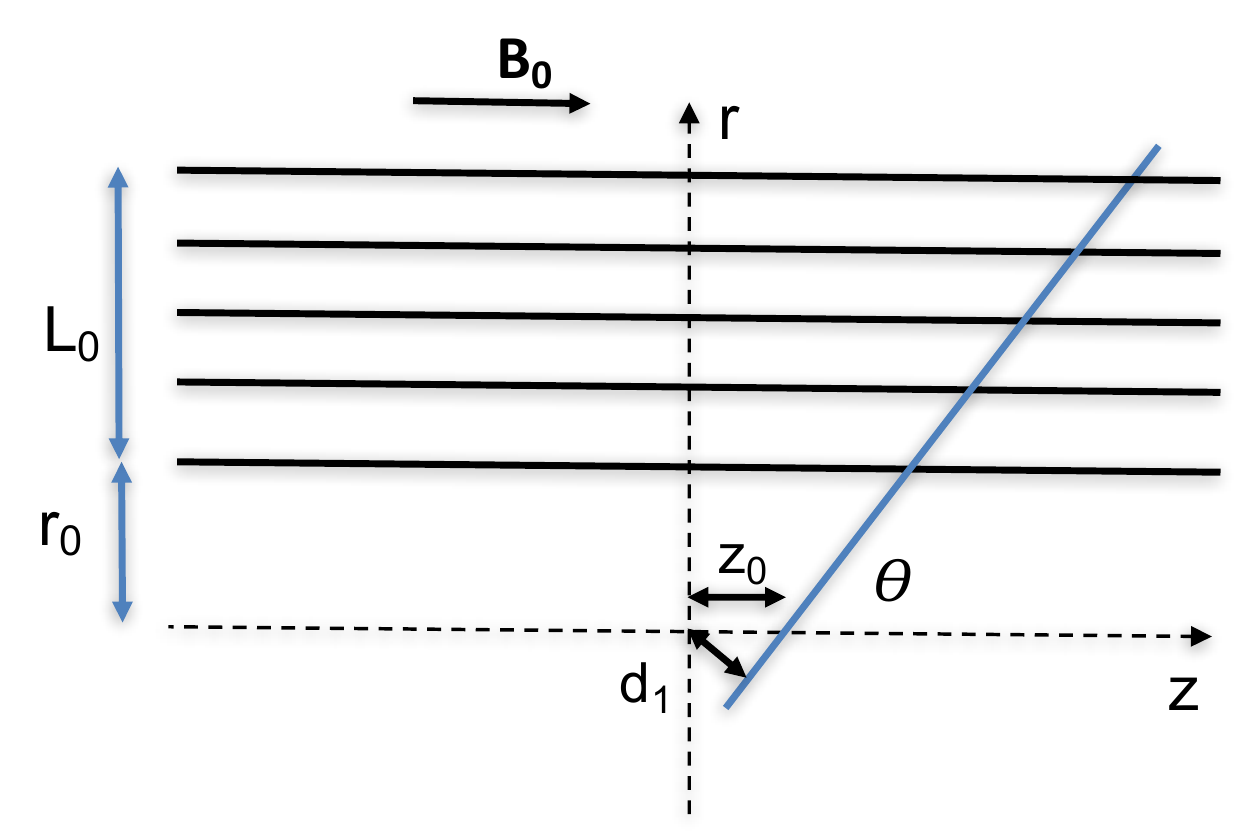}
  \caption{Definition of the track parameters $R, \phi, d_0, \theta,  z_0$. }
  \label{rphiz}
  \end{center}
\end{figure}

\noindent
Finally we present the summary of all results from this report,  applying the derived expressions from the geometries in Fig. \ref{straight_line} and Fig. \ref{parabola} to the  detector geometry from Fig. \ref{rphiz}. The units are $p$\,[GeV/c], $p_T$\,[GeV/c], $L_0$\,[m], $r_0$\,[m], $\sigma_{r\phi}$\,[m], $\sigma_z$\,[m] and $B$\,[T]. The formulas refer to $N+1$ equidistant detector layers of thickness $d/X_0$, where $X_0$ is radiation length of the material.  The total material budget of this arrangement at perpendicular incident angle is therefore $d_{tot}/X_0=(N+1)d/X_0$. We have denoted $\sigma_{r\phi}$ as the position resolution in $r{-}\phi$ direction and $\sigma_z$ as the resolution in $z-$direction. The factor $\beta$ is related to the momentum by $\beta = (p/\sqrt{m^2c^2+p^2)}$. Instead of the angle $\theta$, the pseudorapidity $\eta=-\ln \tan \theta/2$ is used for hadron collisions, so we have $1/\sin \theta = \cosh \eta$ in the following expressions. We define $f(y)=0.0136\,\mathrm{GeV/c}\times \sqrt{y}(1+0.038 \ln y)$.
\\ \\
\subsection{Momentum resolution}
For a track at a given angle $\theta$ the connection between Fig. \ref{parabola} and Fig. \ref{rphiz}a is by $L=L_0/\sin \theta, B=B_0 \sin\theta, p=p_T/\sin \theta$ and the amount of traversed material is increased by the factor $1/\sin \theta$, so by inserting this into Eq. \ref{momentum_resolution_raw} and Eq. \ref{multiple_scattering_final} we have
\bea
     \frac{\Delta p_T}{p_T}\vert_{res.} & = & \frac{\sigma_{r\phi} \, p_T}{0.3\,B_0 L_0^2}\sqrt{\frac{720 N^3}{(N-1)(N+1)(N+2)(N+3)}}  \label{sum_pt_res}  \\
     & \approx &  \frac{12 \, \sigma_{ r\phi} \, p_T}{0.3\,B_0 L_0^2}\sqrt{\frac{5 }{N+5}}    \label{sum_pt_res}  \\
    \frac{\Delta p_T}{p_T}\vert_{m.s.} & = &  \frac{N}{\sqrt{(N+1)(N-1)}} \, \frac{0.0136\,\mathrm{GeV/c}}{0.3 \beta \, B_0 L_0 }\,\sqrt{\frac{d_{tot}}{X_0\,\sin\theta}}\left(1+0.038 \ln \frac{d}{X_0\,\sin\theta}\right) \label{sum_pt_ms} \\
    & \approx &   \frac{0.0136\,\mathrm{GeV/c}}{0.3 \beta \, B_0 L_0 }\,\sqrt{\frac{d_{tot}}{X_0\,\sin\theta}} \label{sum_pt_ms}
\eea
The dependence of momentum resolution on $p_T$ and $\theta$ (or $\eta$) has the general form
\beq
    \frac{\Delta p_T}{p_T} = a\,p_T \oplus \frac{b}{\sin^\frac{1}{2} \theta} \equiv a\,p_T \oplus b \cosh^\frac{1}{2} \eta
\eeq

\subsection{Angular resolution in the $r{-}z$ plane}
The relation between Fig. \ref{straight_line} and Fig. \ref{rphiz}b for Eq. \ref{psi_reso} and Eq. \ref{psi_ms} is $L=L_0/\sin \theta$, $\sigma = \sigma_z \sin \theta$, $p=p_T/\sin \theta$ and the amount of traversed material is increased by the factor $1/\sin \theta$, so we have
\bea
     \Delta \theta\vert_{res.}  & = & \frac{\sigma_z\, \sin^2 \theta }{L_0} \sqrt{\frac{12N}{(N+1)(N+2)}} \label{sum_theta_res} \\
     & \approx & \frac{2\, \sigma_z\, \sin^2 \theta }{L_0} \sqrt{\frac{3}{N+3}} \label{sum_theta_res} \\ 
     \Delta \theta\vert_{m.s.}  & = &\frac{\sin \theta \, }{\beta \, p_T}\,f \left(\frac{d}{X_0\,\sin\theta}\right) \\
    & \approx & \frac{0.0136\,\mathrm{GeV/c}\,\sin \theta \, }{\beta \, p_T}\,\sqrt{\frac{d}{X_0\,\sin\theta} }
     \label{sum_theta_ms} 
\eea
The dependence of angular resolution in the  $r{-}z$ plane on $p_T$ and $\theta$ (or $\eta$) has the general form
\beq
     \Delta \theta = a\,\sin^2 \theta \oplus b\,\frac{\sin^{\frac{1}{2}} \theta}{p_T} \equiv \frac{a}{\cosh^2 \eta} \oplus \frac{b}{p_T \cosh^\frac{1}{2}\eta}
\eeq

\subsection{Angular resolution in $r{-}\phi$ plane}

The relation between Fig. \ref{parabola} and Fig. \ref{rphiz}a for Eq. \ref{phi_angle_reso} and Eq. \ref{phi_angle_ms} is $r=r_0/\sin \theta$, $L=L_0/\sin \theta$, $p=p_T/\sin \theta$ and we have to multiply the result by $1/\sin \theta$ to project the angle onto the $r{-}\phi$ plane, which gives
\bea
       \Delta \phi\vert_{res.}  & = &     \frac{\sqrt{12} \sigma_{r\phi}}{L_0\sqrt{ (N-1)(N+1)(N+2)(N+3)}}
    \sqrt{  
      (16N^3+2N^2-3N)+\frac{60N^3\,r_0}{L_0}+\frac{60N^3\,r_0^2}{L_0^2}
      } \label{sum_phi_res}  \\
        & \approx &  \frac{\sigma_{r\phi}}{L_0}\frac{8\sqrt{3}}{\sqrt{N+5}} \sqrt{1+\frac{15}{4}\frac{r_0}{L_0}+\frac{15}{4}\frac{r_0^2}{L_0^2}}\\ 
       \Delta \phi\vert_{m.s.}  & = & 
    \frac{1}{\beta p_T} \,f \left( \frac{d}{X_0 \sin \theta}\right)\sqrt{
   \frac{N-3/4}{N-1} 
   + \frac{N}{N-1}\left( \frac{r_0}{L_0} \right)
   +  \frac{N^2}{N-1}\left( \frac{r_0}{L_0} \right)^2
   }
   \\
   & \approx &
     \frac{0.0136\,\mathrm{GeV/c}}{\beta p_T} \, \sqrt{ \frac{d}{X_0 \sin \theta} }\sqrt{
     1
   + \left( \frac{r_0}{L_0} \right)
   +  \left( \frac{r_0}{L_0} \right)^2
   }
\label{sum_theta_ms}  
\eea
The dependence of angular resolution in the  $r{-}\phi$ plane on $p_T$ and $\theta$ (or $\eta$) has the general form
\beq
    \Delta \phi = a \oplus \frac{b}{p_T \sin^\frac{1}{2} \theta} = a \oplus \frac{b \cosh^\frac{1}{2} \eta}{p_T}
\eeq

\subsection{Transverse impact parameter resolution}

The relation between Fig. \ref{parabola} and Fig. \ref{rphiz}a in Eq. \ref{delta_d0_res} and Eq. \ref{delta_d0_ms} is $L=L_0/\sin \theta$, $r=r_0/ \sin \theta$, $p=p_T/\sin \theta$ and we get
\bea  
              \Delta d_0\vert_{res.}  & = &   \frac{3 \sigma_{r\phi}}{\sqrt{(N-1)(N+1)(N+2)(N+3)}} \times  \label{sum_d0_res} \\
       &    &   \sqrt{
       \left(N^3-\frac{N}{3}-\frac{2}{3} \right)
       + \frac{4(2N^3-N^2-N)r_0}{L_0}
       + \frac{4(7N^3-N^2-N)r_0^2}{L_0^2}
       +\frac{40N^3 r_0^3}{L_0^3}
       +\frac{20N^3 r_0^4}{L_0^4} } \no \\
          &   \approx &   \frac{3 \sigma_{r\phi}}{\sqrt{N+5}}  \sqrt{
       1
       + \frac{8r_0}{L_0}
       + \frac{28r_0^2}{L_0^2}
       +\frac{40 r_0^3}{L_0^3}
       +\frac{20 r_0^4}{L_0^4} } \\
                   \Delta d_0\vert_{m.s.}  & = & \frac{r_0}{\beta p_T}\,f \left( \frac{d}{X_0 \sin \theta}\right) \sqrt{
         \frac{N-3/4}{N-1} + 
         \frac{N}{2(N-1)} \left( \frac{r_0}{L_0} \right) + 
         \frac{N^2}{4(N-1)} \left( \frac{r_0}{L_0}\right)^2 } \,  \\
         & \approx & \frac{0.0136\,\mathrm{GeV/c}}{\beta p_T}r_0\,\sqrt{ \frac{d}{X_0 \sin \theta}}\sqrt{
        1+ 
         \frac{1}{2} \left( \frac{r_0}{L_0} \right) + 
         \frac{N}{4} \left( \frac{r_0}{L_0}\right)^2 } 
       \label{sum_d0_ms} 
\eea
The dependence of transverse impact parameter resolution  on $p_T$ and $\theta$ (or $\eta$) has the general form
\beq
    \Delta d_0 = a \oplus \frac{b}{p_T \sin^\frac{1}{2} \theta}  = a \oplus \frac{b \cosh^\frac{1}{2} \eta}{p_T}
\eeq

\subsection{Longitudinal impact parameter resolution}

The relation between Fig. \ref{straight_line} and Fig. \ref{rphiz}b in Eq. \ref{delta_z0_res} and Eq. \ref{delta_z0_ms} is $\Delta z_0 = \Delta d_1/\sin \theta$, $L=L_0/\sin \theta$, $r=r_0/\sin \theta$, $p=p_T/\sin \theta$, $\sigma = \sigma_z \sin \theta$ and we get

\bea 
              \Delta z_0\vert_{res.}  & = & \frac{2\sigma_z}{\sqrt{(N+1)(N+2)}}
              \sqrt{
       \left(N+\frac{1}{2}\right)+\frac{3N r_0}{L_0}+\frac{3N r_0^2}{L_0^2} 
       }  \label{sum_z0_res}  \\
                   &   \approx & \frac{2\sigma_z}{\sqrt{N+3}}
              \sqrt{ 1+\frac{3 r_0}{L_0}+\frac{3 r_0^2}{L_0^2} }  \no   \\ 
              \Delta z_0\vert_{m.s.}  & = &   \frac{r_0}{\sin \theta}  
             \frac{1}{\beta p_T}\, f\left( \frac{d}{X_0 \sin \theta} \right)
              \label{sum_z0_ms} \\
                & \approx &  
             \frac{0.0136\,\mathrm{GeV/c}}{\beta p_T}\frac{r_0}{\sin \theta}\, \sqrt{ \frac{d}{X_0 \sin \theta} }
              \label{sum_z0_ms}
\eea
The dependence of longitudinal impact parameter resolution on $p_T$ and $\theta$ (or $\eta$) has the general form
\beq
    \Delta z_0 = a \oplus \frac{b}{p_T \sin^\frac{3}{2} \theta} = a \oplus \frac{b\,\cosh^\frac{3}{2} \eta}{p_T}
\eeq

\clearpage
\newpage

\end{document}